\documentclass[reprint,amsmath,amssymb,aip,jcp]{revtex4-1}

\usepackage{graphicx}% Include figure files
\usepackage{bm}% bold math

\usepackage{amsbsy}
\usepackage{color}

    \newcommand{\beq}{\begin{equation}}
    \newcommand{\eeq}{\end{equation}}
    \newcommand\beqa{\begin{eqnarray}}
    \newcommand\eeqa{\end{eqnarray}}

\begin{document}

%\preprint{}


\title{Communication: Virial coefficients and demixing in highly asymmetric binary additive hard-sphere
mixtures}



\author{Mariano L\'opez de Haro}
\email{malopez@unam.mx}
\homepage{http://xml.cie.unam.mx/xml/tc/ft/mlh/}
\affiliation{Instituto
de Energ\'{\i}as Renovables, Universidad Nacional Aut\'onoma
de M\'exico (U.N.A.M.), Temixco, Morelos 62580, M{e}xico}

\author{Carlos F. Tejero}
\email{cftejero@fis.ucm.es}
\affiliation{Facultad de Ciencias F\'{\i}sicas, Universidad Complutense de
Madrid, E-28040, Madrid, Spain}
\author{Andr\'es Santos}
\email{andres@unex.es}
\homepage{http://www.unex.es/eweb/fisteor/andres/}
\affiliation{Departamento de F\'{\i}sica, Universidad de
Extremadura, Badajoz, E-06071, Spain}

\begin{abstract}
The problem of demixing in a binary fluid mixture of highly asymmetric additive hard spheres is revisited. A comparison is presented between the results derived previously using truncated virial expansions for three finite size ratios with those that one obtains with the same approach in the extreme case in which one of the components consists of point particles. Since this latter system is known not to exhibit fluid-fluid segregation, the similarity observed for the behavior of the critical constants arising in the truncated series in all instances,  while not being conclusive, may cast serious doubts as to the actual existence of a demixing fluid-fluid transition in disparate-sized binary additive hard-sphere mixtures.
\end{abstract}

\date{\today}

%\keywords{}

\maketitle

An analysis of the solution of the Percus--Yevick integral equation for
binary additive hard-sphere (HS) mixtures\cite{LR64} leads to the conclusion
that no phase separation into two fluid phases exists in these
systems. The same conclusion is reached if one considers the most
popular equation of state proposed for additive HS mixtures,
namely the Boubl\'{\i}k--Mansoori--Carnahan--Starling--Leland (BMCSL) equation of state.\cite{B70,MCSL71} For a long time the belief was
that this was a true physical feature. Nevertheless, this belief
started to be seriously questioned after Biben and Hansen\cite{BH91} obtained fluid-fluid segregation in such mixtures
out of the solution of the Ornstein--Zernike equation with the
Rogers--Young closure, {provided the size disparity was large enough}.
{More recently, an accurate equation of state derived by invoking some consistency conditions\cite{S12c} does predict phase separation}.
The importance of this issue resides in the
fact that if fluid-fluid phase separation occurs in additive HS
binary mixtures, it must certainly be entropy driven. In contrast,
in other mixtures such as molecular mixtures, {temperature plays a non-neutral role and demixing is a free-energy driven phase transition.}

The demixing problem has received a lot of attention in the
literature in different contexts and using different approaches.
For instance, Coussaert and Baus\cite{CB97,CB98a,CB98b} have proposed an
equation of state with improved virial behavior for a binary
HS mixture that predicts a fluid-fluid transition at very
high pressures (metastable with respect to a fluid-solid one). On
the other hand, Regnaut \textit{et al.}\cite{RDA01} have examined
the connection between empirical expressions for the contact
values of the pair distribution functions and the existence of
fluid-fluid separation in HS mixtures. Finally, in the
case of highly asymmetric binary additive HS mixtures, the
depletion effect has been invoked as the physical mechanism behind
demixing (see, for instance, Refs.\
\onlinecite{DRE98,DRE99a,DRE99b,AA06,AWRE11}
and the bibliography
therein) and an effective one-component fluid description has been employed.
{It is worth remarking that the entropic forces driving a possible demixing transition become much more important as the dimensionality increases, so that demixing in four- and five-dimensional HS systems is much less elusive.\cite{YSH00} In the limit of infinite dimensionality, demixing  becomes possible even in the presence of negative nonadditivity.\cite{SH05}}

This paper addresses the (in our view) still unsolved and controversial problem of
demixing in {three-dimensional} binary mixtures of additive HS. Our system consists of a binary
fluid mixture of $N=N_1+ N_2$ additive HS of species $1$ and $2$ whose diameters are
$\sigma_1$ and $\sigma_2$, respectively, with $\sigma_1>\sigma_2$, so that the range of the repulsion between particles of species $1$ and $2$ is  $\sigma_{12}=\frac{1}{2}\left(\sigma_1+\sigma_2\right)$. The thermodynamic properties of the mixture can be
described in terms of the number density (which for this system is
given by $\rho\equiv N/V$, with $V$ the volume), the mole fraction
of the big spheres $x\equiv N_1/N$, and the parameter $\gamma\equiv
\sigma_2/\sigma_1$, which measures the size asymmetry. Also convenient for later use is the packing fraction $\eta \equiv (\pi/6) \rho \sigma_1^3[x+(1-x)\gamma^3]$. We will consider as starting point for our analysis the available information on the (in principle exact) virial expansion of the equation of state. In general, one may express the virial coefficients of a binary
HS mixture as
\begin{equation}
\label{5}
B_n(x,\gamma)=\sum_{m=0}^{n}B_{m,n-m}(\gamma)\frac{n!}{m!(n-m)!}x^{m}(1-x)^{n-m},
\end{equation}
where the partial  (composition-independent) virial
coefficients $B_{m,n-m}(\gamma)$ ($m=0,1,\ldots,n$) have been introduced.
Analytical expressions are known for $B_2(x,\gamma)$\cite{K55}
and $B_3(x,\gamma)$,\cite{KM75} while $B_4(x,\gamma)$ and  up to
$B_7(x,\gamma)$ have been evaluated numerically for various size
ratios.\cite{SFG96,EACG92,EAGB98,WSG98,VM03} Recently,
Lab\'{\i}k {and Kolafa have} developed an accurate algorithm to
compute virial coefficients up to $B_8(x,\gamma)$ at a number of
size ratios.\cite{LK13}  The specific values of the partial virial
coefficients $B_{m,n-m}(\gamma)$ with $n=4$--$8$ for $\gamma=0.05$, $0.1$, and $0.2$ were reported in Table 1 of Ref.\ \onlinecite{HML10}. These most recent results, apart from providing
the new eighth virial coefficients, also improve on the numerical values
of the lower ones. A recent review on
virial expansions, including an extensive list of references and a
description of the difficulties associated with the computation of
higher virial coefficients, has been written by Masters.\cite{M08a}

A convenient way to study demixing in binary additive
HS mixtures is to look at the loss of convexity of the
Helmholtz free energy per particle $f\equiv f(\rho,x,\gamma)$. For our binary HS mixture it
reads
\beq
f = f_{{\rm id}}+f_{{\rm ex}},\label{1}
\eeq
with the ideal contribution $f_{{\rm id}}$ given by
\beq
\label{1a} \beta f_{{\rm id}}=-1+
x\ln\left(\rho x\Lambda_1^3\right)+(1-x)\ln\left[\rho(1-x)\Lambda_2^3\right],
\eeq
and, in terms of $B_{n+1}(x,\gamma)$, the excess contribution $f_{{\rm ex}}$ given by
\begin{eqnarray}
\label{1b} \beta f_{{\rm ex}} = \sum_{n=1}^{\infty}\,
\frac{1}{n}\,B_{n+1}(x,\gamma)\rho^n.
\label{fex}
\end{eqnarray}
In the above formulae, $\beta\equiv 1/k_BT$ (where $k_B$ is the Boltzmann constant and $T$ the absolute temperature)  plays only the role of a
scale factor, and $\Lambda_i$ ($i=1,2$) is the thermal de Broglie
wavelength of the particles of species $i$.
In the present thermodynamic representation, where $\rho$ and $x$ are the independent variables,
the condition for the occurrence of a spinodal instability reads
\begin{equation}
\label{2}  \left(\frac{\partial^2 f}{\partial \rho^2}
+\frac{2}{\rho}\frac{\partial f}{\partial \rho}\right)
\frac{\partial^2 f}{\partial x^2} -\left( \frac{\partial^2
f}{\partial \rho
\partial x}\right)^2=0.
\end{equation}

In two instances, namely the limiting cases of a pure HS
system ($\gamma=1$) and that of a binary mixture in which species
$2$ consists of point particles ($\gamma=0$), it is known that there is no fluid-fluid
separation.\cite{V98} For size ratios other than $\gamma=1$ and $\gamma=0$, once
$\gamma$ is fixed, the constants corresponding to the lower critical consolute point, $\rho_c$ and
$x_c$, should be found by determining the minimum of the curve
obtained from the use of Eq.\ (\ref{2}). However, due to the fact that
the virial coefficients beyond the eighth are unknown, the exact
expression for $f(\rho,x,\gamma)$ is also unknown. Hence, either one
truncates the series in Eq.\ (\ref{1b}) after the term with $n=7$ or
uses an approximate compressibility factor $Z_{{\rm app}}(\rho)$ (in
which case $\beta f_{{\rm ex}}\simeq \int_{0}^{\rho} d \rho' [Z_{{\rm
app}}(\rho')-1]/{\rho'}$) to approximate the true Helmholtz
free energy.

As already observed by Vlasov and Masters\cite{VM03} and L\'opez de
Haro and Tejero,\cite{HT04} truncation of a virial series [which
is equivalent to truncating the series in Eq.\ (\ref{1b})] can
produce dramatic effects on the resulting critical behavior of the
mixture. More recently, by working with
the truncated virial series and systematically adding one more known
coefficient from the second to the eighth, L\'opez de Haro \emph{et al.}\cite{HML10}  obtained the
(apparent) critical consolute point for three mixtures of size ratios
$0.05$, $0.1$, and $0.2$. In the three cases it was found that the values of
the critical pressures and packing fractions monotonically increase with
the truncation order. Extrapolation  of these results to infinite order
suggests that the critical pressure diverges to infinity and the
critical packing fraction tends towards its close-packing value, thus
supporting a non-demixing scenario, at least for the three systems
investigated. In Ref.\ \onlinecite{HML10} it was also found that the same trends were obtained
when the unknown exact virial coefficients beyond the eighth one are
estimated from Wheatley's extrapolation formula\cite{W99,BS08} or when the
BMCSL equation of
state\cite{B70,MCSL71} (which does
not predict demixing) is ``amended" by replacing a number of
approximate virial coefficients by the exact ones.  This
shows the extreme sensitivity of the demixing phenomenon to slight
changes in the approximate equation of state that is chosen to describe
the mixture.

A more detailed analysis of the results for the critical pressures $p_c(k)$ obtained by L\'opez de Haro \emph{et al.}\cite{HML10} allows one to get an insight of the behavior of $p_c$
with the truncation order $k$. In fact, a
log-log plot of $p_c(k)$ vs.\ $k$ shows a quasi-linear behavior,
consistent with a power law $p_c(k)\approx  A k^\mu$, with an exponent
$\mu\approx 1.7$--$2$ that slightly depends on the size ratio.

The argument that the truncated virial series are prone to exhibit
demixing, albeit with larger and larger critical pressures, can be
reinforced, as will be discussed in this paper, by analyzing a binary mixture in which species
$2$ consists of point particles, so that $\gamma=0$. In that limit the
exact free energy is $\beta f(\rho,x,\gamma=0)=x \beta
f_\text{pure}(\eta)-(1-x)\ln(1-\eta)$, where $f_\text{pure}$ is the free energy of a pure HS fluid evaluated at the same packing fraction $\eta$ as that of the mixture. Note that in this limiting case the virial coefficients of the mixture are
directly related to the ones of the pure fluid, which are known up to the tenth.\cite{vR93, LKM05,CM05,CM06} Further, and as mentioned previously, this system is known to
lack a demixing transition\cite{V98} but, as shown below, the truncated
virial series exhibits artificial critical points with the same
qualitative features as observed for the mixtures with size ratios $\gamma=0.05$, $0.1$, and $0.2$.

\begin{figure}[ht]
\includegraphics[width=\columnwidth]{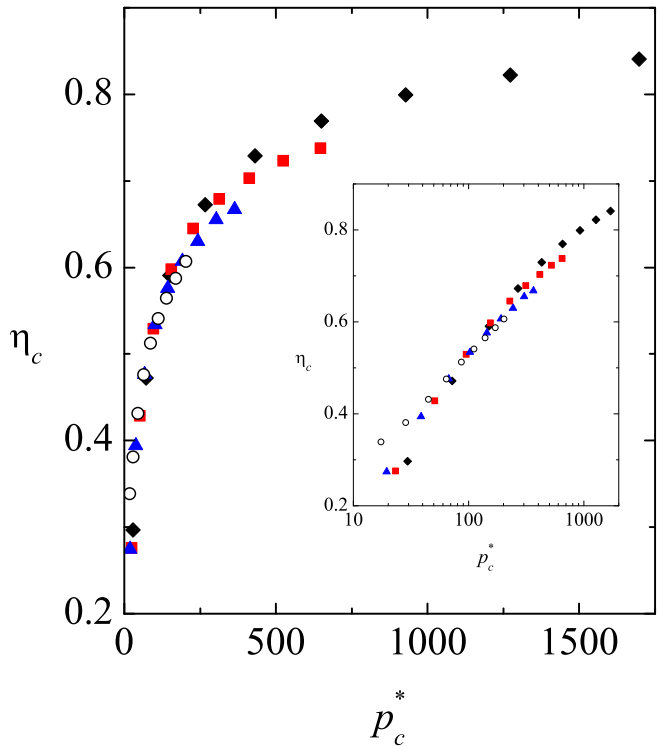}
\caption{Critical packing fraction $\eta_c$ vs.\ reduced critical
pressure $p_c^*$ in binary HS mixtures of different size
ratios $\gamma$ as computed from truncated virial expansions keeping
successively two, three, four, five, six, seven, eight, nine, and ten `exact'
virial coefficients. In the case of non-zero $\gamma$, the `exact' ninth and tenth virial coefficients have been estimated using Wheatley's extrapolation formula. Diamonds: $\gamma=0$; squares: $\gamma=0.05$; triangles:
$\gamma=0.1$; open circles: $\gamma=0.2$. {The inset shows the representation with the critical pressure in logarithmic scale.}}
\label{fig1}
\end{figure}

In Fig.\ \ref{fig1} we illustrate the trends observed with different values of the size ratio both for the
reduced critical pressure $p_c^*\equiv \beta p_c\sigma_1^3$ and packing fraction $\eta_c$ as one adds one more
`exact' virial coefficient (up to the tenth) each step to the truncated virial series. In the case of non-zero $\gamma$, the ninth and tenth virial coefficients have been computed with Wheatley's extrapolation formula.\cite{W99,note_13_02} As already pointed out in Ref.\ \onlinecite{HML10}, for non-zero $\gamma$ one does not know the convergence properties of the virial series and hence whether the demixing transition in such binary mixtures is
either stable, metastable with respect to freezing, or nonexistent cannot be ascertained on the basis of the previous results alone. On the other hand, the absence of the demixing transition is certain\cite{V98} for $\gamma=0$ and the trends observed with the truncated series in this case for the $\eta_c$ vs.\ $p_c^*$ curve are strikingly similar to those that arise for the same curve when $\gamma=0.05$, $0.1$, and $0.2$. Although not shown, if one considers the limit $\gamma=0$ in the BMCSL equation of state, the results obtained from truncating this latter are virtually indistinguishable from the ones shown in Fig.\ \ref{fig1}.

In conclusion, while not settling definitely the matter and contrary to approaches based on either approximate integral equations or on an effective one-component description, the above results provide further evidence that it is {plausible} that a stable demixing fluid-fluid transition {does not} occur in (three-dimensional) additive binary HS mixtures with non-zero size ratio.

\acknowledgments

A.S. acknowledges the financial support of the Spanish Government through Grant No. FIS2010-16587 and  the Junta de Extremadura (Spain) through Grant No.\ GR10158 (partially financed by FEDER funds).

%\bibliographystyle{apsrev}

%\bibliography{D:/Dropbox/Public/bib_files/liquid}

\end{document}